\documentclass[twocolumn,showpacs,preprintnumber,amsmath,amssymb,aps]{revtex4}
\usepackage{graphicx}
\begin{document}

\title{Stripes, Clusters, and Nonequilibrium Ordering for Bidisperse 
Colloids with Repulsive Interactions} 
\author{C. Reichhardt and C.J. Olson Reichhardt} 
\affiliation{ 
Theoretical Division and Center for Nonlinear Studies,
Los Alamos National Laboratory, Los Alamos, New Mexico 87545}

\date{\today}
\begin{abstract}
We show that two-dimensional bidisperse assemblies of colloids with strictly 
repulsive interactions 
exhibit stripe, cluster, and partially crystallized states 
when driven over a quenched random substrate. 
The nonequilibrium states on a substrate are significantly more ordered than 
equilibrium states both with and without substrates.
A minimum substrate strength is necessary to induce the
nonequilibrium pattern formation.
Our results suggest that a combination of driving and quenched
disorder offers a new approach to controlling 
pattern formation in colloid mixtures. 
\end{abstract}
\maketitle

\vskip2pc

Understanding how assemblies of particles organize  
is an outstanding problem in both equilibrium and nonequilibrium systems. 
The ability to control 
pattern formation
would have a profound impact on
applications that use large-scale self-assembly    
to create specific pattern morphologies.  
In equilibrium, it is known that particles with competing repulsive and
attractive interactions
can organize 
into clusters, stripes, and spongelike textures \cite{Seul}. 
Similar patterns can also arise
for particles with strictly repulsive potentials 
that have a two step form \cite{Stripe1}.
Recently, it was shown in both simulations 
and experiments that a two-dimensional (2D)
bidisperse assembly of colloids interacting via repulsive magnetic dipoles of
two different strengths
can cluster into spongelike patterns \cite{Maret}.  
This suggests that other types of ordering are possible for
bidisperse repulsively interacting particles.

Here, we demonstrate that a variety of distinct partially 
ordered states can occur for a bidisperse 
system of strictly repulsively interacting 
colloids driven over a random substrate that 
is sufficiently strong.
There has been extensive work on the ordering of monodisperse
charged particles moving over random quenched disorder 
for systems such as vortices in type-II superconductors 
\cite{Giamarhi,Koshelev,Pardo},
moving Wigner crystals \cite{Jensen} and colloids \cite{Chen}. 
In the absence of quenched disorder, a monodisperse system forms a
triangular lattice, while strong quenched disorder distorts the lattice and
generates numerous topological defects.  These defects can be partially
annihilated and the order partially regained by driving the system into a 
nonequilibrium moving smectic state
\cite{Giamarhi,Koshelev,Pardo,Jensen,Chen}.   
It is not known what type of nonequilibrium configurations 
would occur for {\it two} species of particles  
driven with identical force
over quenched disorder. In contrast to the monodisperse case,
for a bidisperse system even the equilibrium states in the absence of 
a random substrate are intrinsically disordered \cite{Maret}, so 
one might expect the addition of quenched disorder 
to further disorder the system. 
Instead, we find that for certain regimes, 
nonequilibrium bidisperse colloid assemblies
moving over quenched disorder have {\it more} topological order
than the corresponding equilibrium states, 
even those {\it without} quenched disorder. 
The quenched disorder must be sufficiently strong 
for the topological ordering to occur. 
We argue that for strong quenched disorder,
significant plastic deformations of the driven colloid configuration 
occur due to the fact that
the different species move at different average velocities
close to the depinning threshold.
We show that the system can organize into stripes with triangular ordering
within each stripe.
When the difference between the two species is 
small, a moving smectic state forms, whereas when the disparity is very large  
moving cluster states form. 
Our results are robust for a wide range of colloid densities, 
quenched disorder strengths, and system sizes.           

We employ Brownian dynamics to simulate a 2D system of size 
$L \times L$
containing $N$ colloids and $N_{p}$ potential traps
with periodic boundary conditions in the
$x$ and $y$ directions. 
The overall density of the system is 
$n=N/L^2$.
The colloids interact via a screened Coulomb potential 
$V(R_{ij}) = 
(E_{0}/R_{ij})\exp(-\kappa R_{ij})$,
where ${\bf R}_{i(j)}$ is the position of colloid $i(j)$,
$R_{ij}=|{\bf R}_{i}-{\bf R}_{j}|$,
$E_{0}= Z^{*2}/(4\pi\epsilon\epsilon_{0}a_0)$, 
$Z^*$ is the unit of charge, $\epsilon$ is the solvent dielectric 
constant, and $1/\kappa$ is the screening length.
The unit of distance in the simulation is $a_0$,
and unless otherwise noted, 
$L=48a_0$.
In this work we fix $\kappa = 4a_0$, 
which is reasonable for experiments on colloids in nonpolar
fluids \cite{Hsu}.
Forces are measured in units of $F_0=E_{0}/a_0$.
The dynamics of colloid $i$ are given by the 
equation of motion  
\begin{equation}
\eta\frac{ d{\bf R}_{i}}{dt} = -q_{i}\sum^{N}_{j = 1}q_j\nabla V(R_{ij}) -  
\sum^{N_{p}}_{k = 1}\nabla V_{p}(R_{ik}) + {\bf F}_{D} + {\bf F}_i^T  
\end{equation}
where $\eta$ is the damping term, $q_{i(j)}$ is the dimensionless charge
of colloid $i(j)$, $R_{ik}=|{\bf R}_{i}-{\bf R}_{k}|$,
and ${\bf R}_k$ is the position of trap $k$.
In this model, hydrodynamic effects are neglected; 
such effects can be strongly screened
in a system confined within 2D walls.
To introduce bidispersity to the system,
half of the colloids have charge $q_i=q_A$ and the other half have
$q_i=q_B$, where we fix $q_B=1$.  
The quenched random substrate is modeled by randomly distributed parabolic
traps of density $n_p$ and 
radius $r_p=0.1a_0$ with $V_p(R_{ik})=-(F_p/2r_p)(R_{ik}-r_p)^2$ for
$R_{ik} \le r_p$ and zero interaction for $R_{ik} > r_p$.
Here $F_p$ is the maximum pinning force.
This model for quenched random disorder has given results comparable with other
models in monodisperse particle systems.
The externally applied driving force 
is identical for all particles and is given by ${\bf F}_D=F_D{\bf {\hat x}}$.
A uniform drive of this type 
could be created electrophoretically \cite{Electro}.
Thermal effects are modeled by random Langevin kicks with 
$\langle F^{T}_{i}\rangle = 0$ and 
$\langle F^{T}_{i}(t)F^{T}_{i}(t^{\prime})\rangle = 
2\eta k_{B}T \delta_{ij}\delta(t - t^{\prime})$.
The initial colloid configurations are obtained using simulated annealing.
We then set $F^T=0$ and gradually increase $F_D$
in increments of $\delta F_D=0.001$ every $\tau=5000$ simulation time steps. 
After each drive increment, 
once the system has reached a stationary state 
we measure
the average velocity 
$\langle V_x\rangle=\left\langle (1/N)\sum_{i=1}^{N}{\bf v}_i \cdot {\bf {\hat x}}\right\rangle$,
where ${\bf v}_i$ is the velocity of colloid $i$.
The depinning threshold $F_C$ corresponds to the value of $F_D$
at which $\langle V_x\rangle=0.04F_0$.
As an example, 
for $a_{0} = 0.6\mu m$, $\epsilon = 2$, and $Z^* = 300e$,
$F_{0} = 27.8$ pN. 

\begin{figure}
\includegraphics[width=3.3in]{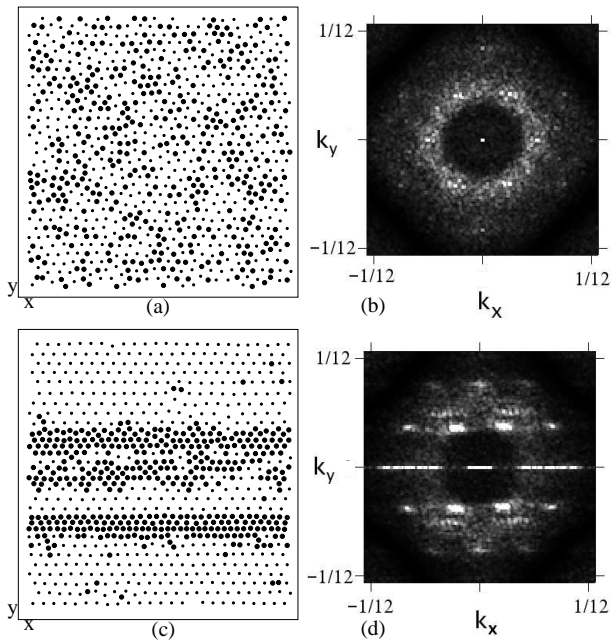}
\caption{
A bidisperse system of colloids with $q_A/q_B=3$ and $n=0.385/a_0^2$.
(a,c) Real space colloid configurations.  Small circles: species A;
large circles: species B. 
(b,d) The corresponding structure factor $S({\bf k})$.
(a,b) Equilibrium state with no pinning
or driving. (c,d) Moving stripe state for $F_{p} = 1.0$, $n_p=4.0/a_0^2$, and 
$F_{D}/F_{C} = 3.0$.
}
\label{fig:image}
\end{figure}

We first consider a system where the ratio between the charges of
the two colloid species is $q_{A}/q_{B} = 3$. 
In the absence of quenched disorder, the colloids form a 
disordered mixed assembly after annealing, illustrated in 
Fig.~\ref{fig:image}(a).
To characterize the configuration, we determine the
structure factor,
$S({\bf k})=(1/N)\left|\sum_{i= 1}^{N}\exp(-i{\bf k}\cdot {\bf R}_i)\right|^2$. 
In Fig.~\ref{fig:image}(b), $S({\bf k})$ for
the equilibrium case with $N=864$ colloids
at density $n=0.385/a_0^2$ shows a ring structure 
which is characteristic of disordered systems.
We also analyze the fraction of six-fold coordinated colloids $P_{6}$ obtained
from a Voronoi construction, $P_6=(1/N)\sum_{i=1}^{N}\delta(z_i-6)$, 
where $z_i$ is the coordination number of colloid $i$.
In a triangular lattice
all the colloids have $z_i=6$, giving $P_{6} = 1.0$. 
For the configuration shown in Fig.~\ref{fig:image}(a),
$P_{6}=0.41$, indicating that a large fraction of the colloids
have $z_i\ne 6$.  
If we anneal in the presence of quenched
disorder, we find a similar disordered state as 
measured by $S({\bf k})$ and $P_{6}$. In general,
for stronger quenched disorder the system becomes more disordered, 
producing increased smearing 
in the ring structure of $S({\bf k})$ 
and reducing $P_{6}$ to $P_6=0.3$.      

\begin{figure}
\includegraphics[width=3.3in]{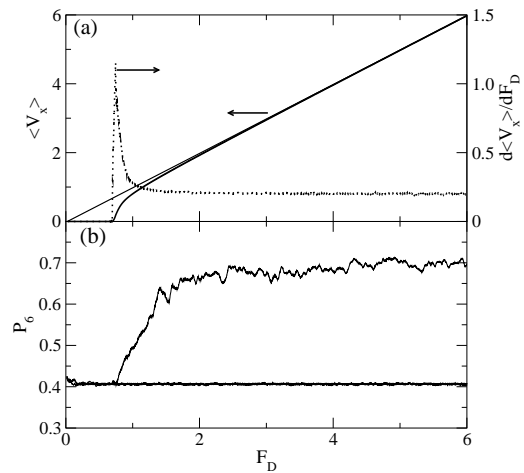}
\caption{
(a) Average velocity $\langle V_{x}\rangle$ vs applied drive $F_{D}$ for 
the system in Fig.~1.  Light upper line: sample with $F_p=0$; dark lower
line: sample with $F_p=1.0$; dashed line: 
$d\langle V_{x}\rangle/dF_{D}$ for the $F_p=1.0$ case.
(b) 
$P_6$ vs $F_D$ for
the same system.  Upper curve: $F_{p} = 1.0$; lower curve: $F_{p} = 0.$  
}
\label{fig:velocity}
\end{figure}

In Fig.~\ref{fig:velocity}(a) we plot the dynamic response of the system 
given by the average colloid velocity
$\langle V_{x}\rangle$ versus external drive $F_{D}$
for 
$F_p=0$ (light line), and
for a system with traps of strength $F_p=1.0$ and 
density $n_p=4.0/a_0^2$ (heavy line).
For $F_{p} = 0$, the velocity response is strictly linear.  
In the presence of traps, there is a 
clear depinning threshold $F_C=0.7$ for colloid motion
followed by a nonlinear 
regime which crosses over to an ohmic regime at higher $F_{D}$.
For $F_p=1$, Fig.~\ref{fig:velocity}(a) shows that
$d\langle V_x\rangle/dF_D$ 
has a peak in the
nonlinear regime at $F_D=0.75$ and flattens
in the ohmic region for $F_{D} > 2.0$. 
These features in the velocity-force curves are the same as 
those previously observed for monodisperse
systems with quenched disorder \cite{Koshelev,Jensen,Chen}, where
the onset of the ohmic response was correlated with a dynamical
reordering of the system into a moving smectic state.

In Fig.~\ref{fig:velocity}(b) we plot
the corresponding $P_{6}$ versus $F_{D}$ curves for the systems
with and without quenched disorder. 
For $F_p=0$, $P_6=0.41$ for all $F_D$.
For $F_p=1.0$, we still find $P_6=0.41$ in the pinned region
at $F_D/F_C<1$. 
Above the depinning transition, $P_{6}$ increases over the
same range of $F_D$ where there is a nonlinear response
in the velocity force curves, as seen in Fig.~\ref{fig:velocity}(a).
Here, the system moves plastically 
with portions of the colloids remaining temporarily trapped while
other colloids move around the trapped colloids.
For $F_{D} > 2.0$,
$P_{6}$ saturates at $P_6\approx 0.7$. 
This result indicates
that the moving system with quenched disorder 
is {\it more ordered} than the equilibrium system without 
quenched disorder. 
To illustrate the nature of this order,
in Fig.~\ref{fig:image}(c) 
we show 
the colloid configuration
for the system with $F_p=1.0$ at 
$F_D/F_C=3.0$, and we plot the
corresponding $S({\bf k})$ in Fig.~\ref{fig:image}(d).
Here  a stripe ordering occurs where the  
species have partially segregated. 
Complete segregation is prevented since this would produce a strong
charge inhomogeneity in the system.
Local triangular order appears
within each stripe.
The structure factor has a smecticlike form, indicative of the presence
of stripes, as well as second order peaks at 
smaller scales resulting from the
in-stripe triangular ordering.

\begin{figure}
\includegraphics[width=3.25in]{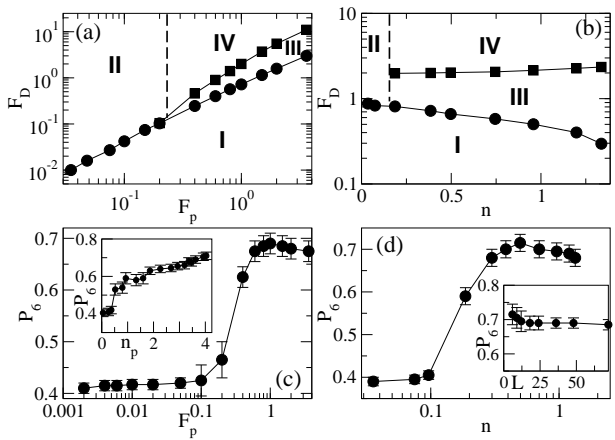}
\caption{
(a) The different regimes for the system in Fig.~1 
at $n=0.385/a_0^2$ and $n_p=4.0/a_0^2$
as a function of
$F_D$ and $F_p$.  I: pinned regime; II: 
moving disordered regime;
III: plastic flow regime; IV: moving stripe regime, as in
Fig.~1(c).
Circles: the depinning threshold $F_{C}$.
Squares: stripe formation threshold.
(b) The different regimes for $F_{D}$ vs density $n$ 
at fixed $F_{p} = 1.0$ and $n_p=4.0/a_0^2$. 
Symbols are the same as in (a).
(c) $P_{6}$ versus $F_{p}$ for $n=0.385/a_0^2$ and $n_p=4.0/a_0^2$
at 
$F_{D}/F_{C} = 3.0$. 
Inset: $P_{6}$ versus pinning density $n_{p}$ for fixed $F_{p} = 1.0$
and $n = 0.385/a_0^2$ at $F_D/F_C=3.0$. 
(d) $P_{6}$ versus 
$n$ for fixed $F_p=1.0$ and $n_p=4.0/a_0^2$
at 
$F_{D}/F_{C} = 3.0$. 
Inset: $P_{6}$ versus system size $L$ for fixed
$F_{p} = 1.0$, $n=0.385/a_0^2$, $n_p=4.0/a_0^2$, and 
$F_D/F_C=3.0$.    
}
\label{fig:phases}
\end{figure}

We next show that sufficiently strong quenched disorder must be 
present for the stripe patterns to occur. 
In Fig.~\ref{fig:phases}(a) 
we outline the four distinct regimes that we find as a
function of driving force $F_D$ and pinning force $F_p$ for a system
with $n=0.385/a_0^2$ and $n_p=4.0/a_0^2$.
At all $F_p$, the colloids are pinned in region I for $F_D<F_C$. 
As $F_D$ is increased above $F_C$,
we find that for $F_p<0.4$, the system never orders into a stripe state.
Instead, the disordered mixture depins with little or no plastic deformation,
and the moving state, marked region II in Fig.~\ref{fig:phases}(a),
retains the same disordered configuration as the
pinned state.  We have tested different values
of $\delta F_D$ and find no change in the location of region II.
For $F_p \ge 0.4$, a plastic flow state (region III) appears above depinning.
Here the colloids
are repeatedly trapped and escape from the traps but stripes do not
completely form. 
Finally, for high enough values of $F_p$ and $F_D$ we find region IV, the
moving stripe state.
In Fig.~\ref{fig:phases}(c) we plot $P_{6}$ as a function of $F_p$ 
for the same system in Fig.~\ref{fig:phases}(a) at fixed 
$F_{D}/F_{C} = 3.0$.  
For $F_{p} > 0.4$, $P_{6}$ saturates near $0.7$ 
and the moving stripe state forms. 
In contrast, for $F_{p} < 0.4$ the moving state has $P_{6} = 0.41$, 
the same value measured in the pinned state. 
This indicates that there is a critical disorder 
strength that is necessary for the formation of 
the stripe state to occur. 

In Fig.~\ref{fig:phases}(b) we plot the locations of the four regimes
as a function of $F_D$ versus colloid density $n$ for fixed $F_p=1.0$ and
$n_p=4.0/a_0^2$.
The depinning threshold $F_C$ marking the end of the pinned region I
drops to lower values of $F_D$ as $n$ increases since the colloid-colloid
interactions become stronger relative to $F_p$, making the traps less
effective at pinning the colloids.
A reordering into the moving stripe state (region IV) occurs only
for $n\ge 0.2/a_0^2$; for smaller values of $n$, we find the moving disordered
region II at high drives.
This is confirmed by
Fig.~\ref{fig:phases}(d), which shows $P_6$ versus $n$ for the same system
in Fig.~\ref{fig:phases}(b) at fixed $F_D/F_C=3.0$.
For $n<0.2/a_0^2$ in the moving region, $P_6 \sim 0.4$.
Here the system does not reorder since at low densities the colloids
are far enough apart that they interact only weakly
and remain in a disordered state.
The pinning density $n_p$ also plays a role in determining whether the moving
stripe state can form.
In the inset of Fig.~\ref{fig:phases}(c) 
we plot $P_{6}$ versus $n_{p}$ for a system
with fixed $F_p=1.0$ and $n=0.385/a_0^2$ at $F_D/F_C=3.0$.
For $n_{p} < 0.5/a_0^2$, $P_6$ remains low, indicating that 
the system does not reorder into the stripe state.
This result shows that a critical 
amount of quenched disorder, as well as a critical strength of disorder, 
is necessary for the moving
system to form the stripe state. 
To show that there are no finite size effects 
for the appearance of the four regions, 
in the inset of Fig.~\ref{fig:phases}(d) we plot $P_{6}$ vs 
system size $L$ at $F_D/F_C=3.0$ in the moving stripe state.  There is a slight
increase in $P_6$ for the smallest values of $L$ since the
small systems form fewer stripes, resulting in a lower number of topological
defects at the interfaces between any two stripes.  
We find that $P_6$ saturates for $L>20$, well
below the size studied throughout this work.

The moving stripe state forms when the 
quenched disorder is strong enough to induce
plastic deformations in the colloid configuration.
The effective pinning force from the substrate decreases for increasing
colloid-colloid interactions in a monodisperse system.
In a bidisperse system, local clustering of colloid species occurs,
as seen in experiments \cite{Maret}. 
Clusters of the weakly charged colloid species B experience a stronger
effective pinning force than clusters of the more strongly charged
colloid species A.  If the quenched disorder is sufficiently strong and
dense, the two species will move at different average velocities {\it even 
though they both couple to the external drive in the same way.}
Stripes form as the faster moving species begins to collect together.
In a co-moving reference frame, this state resembles a system of bidisperse
colloids moving in opposite directions, where a laning instability has been
shown to occur \cite{Lowen}.  In Ref.~\cite{Lowen}, the two species were
driven in opposite directions with no quenched disorder; here, the disparity
in velocity between the two species is induced {\it entirely 
by the quenched disorder}.
If the quenched disorder is weak, the colloid configuration moves elastically
and no velocity difference between the two species occurs, so that no stripes
can form.
We have verified that the average velocity 
distribution for the two different species (not shown) is different 
in region III but the same in region II. 
The stripe phases show less ordering than would occur in a system with only
a single species since topological defects form at the
boundaries between the phases.

\begin{figure}
\includegraphics[width=3.25in]{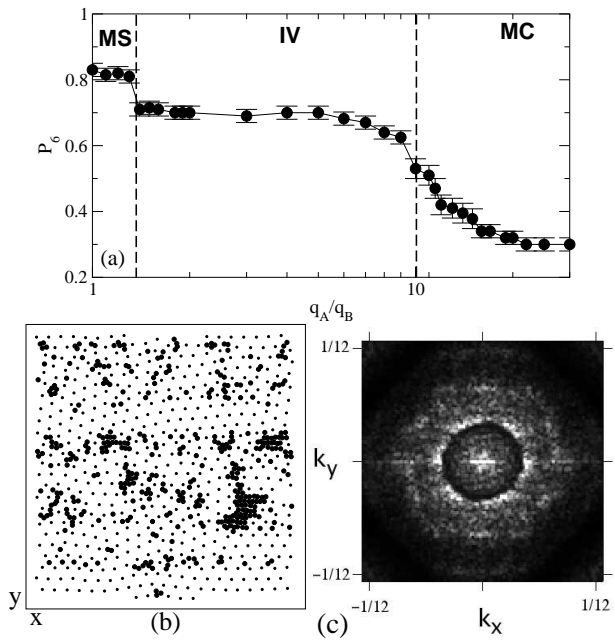}
\caption{
(a) $P_{6}$ versus $q_{A}/q_{B}$ for 
$F_{p} = 1.0$, $n = 0.385/a_0^2$, $n_p=4.0/a_0^2$, and $F_{D}/F_{C} = 3.0$. 
MS: moving smectic region; IV: moving stripe (region IV); MC: moving
clump region.
(b) Real space colloid configuration for the moving clump state at 
$q_{A}/q_{B} = 11$.  Small circles: species A; large circles: species B.
(c) The corresponding structure factor $S({\bf k})$.  
}
\label{fig:clump}
\end{figure}

We now address the effect of the polydipersity on the moving regimes
by varying $q_{A}/q_{B}$.  
A system with 
the same parameters as in Fig.~\ref{fig:velocity}(a) 
and $F_{p} = 1.0$ acts like a monodisperse sample
for $1 <  q_{A}/q_{B} < 1.3$, where it 
reorders into a moving smectic (MS) state 
as observed in previous simulations \cite{Koshelev,Jensen,Chen}.  
For $1.4 < q_{A}/q_{B} \lesssim 10$ the system forms a moving stripe state
(region IV), and for 
$q_{A}/q_{B} \gtrsim 10$ we observe a moving clump (MC) state.
The onset of these different states appears in 
Fig.~\ref{fig:clump}(a) where we plot $P_6$ versus $q_A/q_B$
at $F_D/F_C=3$.
In the moving smectic state, $P_{6} \approx 0.82$,
in the moving stripe state 
$P_{6} \approx 0.7$, and in the moving clump state 
$P_{6}$ is sharply reduced.   
The real space configuration of the moving clump state is illustrated
in Fig.~\ref{fig:clump}(b), and the corresponding structure factor
$S({\bf k})$ in Fig.~\ref{fig:clump}(c) has a multiple ring structure
characteristic of a disordered system containing more than one length
scale.
Here, the different length scales are associated
with the triangular ordering of the higher charge $q_A$ species 
and the interclump distance.
Similar to the moving stripe state, the moving clump state requires sufficient
quenched disorder to form.  As $q_A/q_B$ increases, the size of the individual
clumps decreases.

In summary, we have shown that a bidisperse system of repulsively interacting
colloids driven over quenched disorder can form moving smectic, moving
stripe, and moving clump states.
The moving states are in general 
{\it more} ordered than the equilibrium states that form in the absence
of quenched disorder, in contrast to a monodisperse system.
The stripe state arises when the disorder is strong enough
to induce different average velocities for the two species, leading to
a laning instability similar to that seen for particles driven in
opposite directions.  In our system, the velocity difference is due 
entirely to the quenched disorder, and not to different drives on the
different colloids.
Physical systems in which this model could be realized 
include driven bidisperse colloidal assemblies with charge or 
magnetic interactions driven over optical pinning or a rough wall, 
electron systems with coexisting single and multiple electron 
bubble states, and superconductors with mixtures
of Abrikosov and Josephson vortices. 

This work was carried out under the auspices of the 
%National Nuclear Security Administration 
NNSA of the U.S. 
%Department of Energy 
DoE at 
%Los Alamos National Laboratory 
LANL under Contract No. DE-AC52-06NA25396.

\end{document}